\documentclass[3p,number,final]{elsarticle}
\usepackage[figuresright]{rotating}
\usepackage{afterpage,blindtext,graphicx,booktabs,tablefootnote,hyperref}
\usepackage{enumitem}
\usepackage[flushleft]{threeparttable}

\journal{Pervasive and Mobile Computing}

\begin{document}

\begin{frontmatter}

\title{Does Network Quality Matter? A Field Study of Mobile User Satisfaction}

\author[aalto]{B. Finley\corref{cor1}}
\ead{benjamin.finley@aalto.fi}

\cortext[cor1]{Corresponding author}

\author[aalto]{E. Boz}
\ead{eren.boz@aalto.fi}
\author[aalto]{K. Kilkki}
\ead{kalevi.kilkki@aalto.fi}
\author[aalto]{J. Manner}
\ead{jukka.manner@aalto.fi}
\author[aalto]{A. Oulasvirta}
\ead{antti.oulasvirta@aalto.fi}
\author[aalto]{H. H{\"a}mm{\"a}inen}
\ead{heikki.hammainen@aalto.fi}

\address[aalto]{Department of Communications and Networking, Aalto University, Otakaari 5, Espoo, Finland}

\begin{abstract}
Mobile quality of experience and user satisfaction are growing research topics. However, the relationship between a user’s satisfaction with network quality and the networks real performance in the field remains unexplored.

This paper is the first to study both network and non-network predictors of user satisfaction in the wild. We report findings from a large sample (2,224 users over 12 months) combining both questionnaires and network measurements. We found that minimum download goodput and device type predict satisfaction with network availability. Whereas for network speed, only download factors predicted satisfaction. We observe that users integrate over many measurements and exhibit a known peak-end effect in their ratings. These results can inform modeling efforts in quality of experience and user satisfaction.
\end{abstract}


\end{frontmatter}


\section{Introduction}\label{sec:introduction}
The pervasiveness of mobile network connected devices suggests that understanding the effects of mobile network performance on user quality of experience (QoE) is an important area of research. QoE related results find many applications in, for example, mobile network management and optimization \cite{schatz2014}.

Thus, unsurprisingly, large volumes of research exploring the relationship between QoE and mobile network performance has been published \cite{mitra2014}. Most of these studies have tended to focus on specific use cases such as QoE of mobile video \cite{alreshoodi2013} or specific mobile applications \cite{chen2006,schatz2011,casas2015a,casas2015b}. Relatively few studies have examined the related area of user satisfaction with mobile network concepts such as speed or availability \cite{ozer2013}. Furthermore, studies that have examined this area have been customer satisfaction studies based only on user questionnaires without network data \cite{ozer2013,kuo2009,turel2006}. Additionally, related studies have also typically considered only a small set of network performance features such as download goodput, jitter, and latency and have considered non-network features (such as device type or device quality) out of scope (sometimes because non-network features were unavailable) \cite{chen2006,schatz2011,casas2015a,casas2015b}. Certain non-network features are important to include because they can effect the perception of network quality since, for example, users might incorrectly attribute device related application performance issues to the network.

Relatedly, mobile pervasiveness has also led to an increase in the number of users actively measuring the mobile network to ensure adequate network performance. Evidence can be seen in the popularization of a variety of mobile network measurement apps \cite{goel2015}. However the effect these user measurements and, as mentioned, real world network performance in general have on subjective user satisfaction remains unexplored.

Therefore, in this article we study the network and non-network predictors of user satisfaction with network speed and availability in the context of user device based network measurements. Specifically, we combine user questionnaire responses and those questionnaire respondents empirical mobile network measurements from the device based network measurement platform Netradar \cite{sonntag2013a}. This combination allows us to determine the significant predictors (aka features) of network satisfaction for those end users that themselves actively measure the network and observe the reported results\footnote{Here we emphasize that we are studying users that observe the network measurement results including upload/download goodput and latency through mobile network measurements. We further discuss this issue in Section \ref{discussion}.}. Statistically, we utilize ordinal logistic regression modeling of the questionnaire responses to identify these significant predictors.

Furthermore, our use of the crowd-source based Netradar platform provides both a relatively large sample size (2,224 Finland-based users) and real world data. For studying general network concepts recreating the diversity of real world network conditions in the lab is difficult. Thus implying that field studies with real world data should complement typical QoE and related lab studies. Additionally, for example, lab studies are often short in nature (minutes to hours) and thus might miss longer term (days to months) temporal phenomena such as user adaptation effects \cite{weiss2014}. Though, importantly not all results will differ between lab and field trials and potential differences will depend on the specific use case as Schatz and Eggers \cite{schatz2011} demonstrated in a study combining lab and field trials.

In terms of results, we find that for network availability, minimum download goodput (over a user's measurements), number of frequently measured locations, network operator, and device type (smartphone or tablet) are significant predictors (in our ordinal logistic regression model). Whereas for network speed, we find that minimum, median, and most recent download goodput (over a user's measurements) are highly significant predictors. These network speed satisfaction predictors suggest that both an integration over measurements and a measurement peak-end effect influence the users evaluation. In addition, we find that predictors such as upload goodput, latency (RTT), and device quality, are not significant or only weakly significant given the other predictors. Finally, the overall fits of the result models are only basic thus other unaccounted for factors also likely play a part. Overall, our results have implications for mobile operators in terms of predicting user satisfaction especially in the context of users that measure and observe their own network performance.

We briefly describe the structure of the remainder of the article. Section \ref{NetradarDesc} details the Netradar measurement platform and Section \ref{QoEBackground} details theoretical issues related to QoE and retrospective user evaluations. Section \ref{questionnaire} introduces the questionnaire itself and Section \ref{data_mapping_filtering} details the mapping of questionnaire respondents to measurement data and the filtering of respondents. Section \ref{features} describes the measurement feature extraction process for each respondent. Section \ref{analysis} gives a statistical overview of the resulting features and presents ordinal logistic models for each questionnaire question including significant features. Finally Section \ref{discussion} discusses interpretation issues, Section \ref{related_work} presents related work, and Section \ref{conclusions} gives conclusions.

\section{Netradar Measurement Platform}\label{NetradarDesc}
This section briefly describes the mobile measurement platform Netradar.

Netradar is a popular client-server network measurement platform developed by researchers from Aalto University \cite{sonntag2013a} and initially launched in February 2013. The platform consists of a suite of mobile applications (for different mobile platforms) and associated measurement servers distributed on several different continents. The application sends and receives bulk data to and from the measurement server to estimate network properties such as TCP goodput and round trip time (RTT). The application also simultaneously collects a variety of device information including location, mobile network operator, and platform (Android, iOS, or Windows Phone).

The Netradar client by default performs measurements on demand, in other words, whenever directed to measure by the user selecting the start button in the client user interface. However, the client can also be configured such that measurements are performed in the background (without the need for user intervention) at fixed or random intervals. In the context of the current study, we term measurements that are initiated by the user in the client user interface as \textit{user-initiated measurements}.

During any single network measurement, any part of the Netradar client measurement process (RTT test, TCP test, etc.) might fail for a variety of reasons. For example, the mobile network might not be available (no signal) or the network might be highly congested. These failed measurements might be important indicators of poor network conditions and are not discarded. Even in the case where no network is available, the failed measurement information is stored locally and uploaded when a network connection is available. Thus we can track the total number of failed measurements and the reasons for the failures. In the context of the current study, we term measurements in which no part of the process has failed as \textit{valid measurements} and measurements in which at least one part of the process has failed as \textit{invalid measurements}.

In comparison to similar network measurement platforms Netradar supports all common features (upload/download/latency estimation, coverage maps, etc.) and thus is relatively comprehensive (refer to Table 2 in \cite{goel2015}). In technical terms, both Netradar and almost all other platforms utilize the  bulk transfer capacity method as opposed to other methods such as trains of packet-pair \cite{mikkelsen2015}. Though Netradar uses TCP based upload and download tests whereas similar platforms use HTTP. However the effect of this difference should be minimal. Furthermore, Netradar supports eight different mobile platforms\footnote{Though as noted later, in this work we only consider three platforms due to time constraints with implementing the questionnaire in all platforms.} compared to typical measurement platforms that support only two. 

\section{Relation to QoE and Background Theory}\label{QoEBackground}
This section first positions our study with relation to classical QoE studies and then provides additional background theory in the area of retrospective user evaluations.

\subsection{Relation to QoE}
QoE is a common term in mobile research yet the definition of QoE has been difficult to pin down as many formulations have appeared over the years \cite{mitra2014}. Though recent work, such as the Qualinet project, has made progress in creating a intuitive yet relatively complete definition \cite{callet2013}.

In our current study we take a relatively broad view of QoE that accommodates experiencing over longer time frames and with non-Mean Opinion Score (MoS) scales. In other words, we accommodate two major differences compared to traditional QoE experiments (which comprise a large majority of published work). First our time frame (in months and years) is much longer than traditional lab based QoE experiment time frames (in minutes or hours) and thus encompasses many more episodes. This difference is related to the temporal differences in user retrospective evaluations discussed in Section \ref{RetroQualityEval}. Second we utilize a Likert agree-disagree scale with custom questions rather than the traditional absolute category rating (and corresponding MoS). 

The main reasons for utilizing the Likert method are twofold. Likert scales are well known among the general population and thus do not require significant explanation like the absolute category rating. Such an explanation process would be infeasible in a short crowd-sourced mobile questionnaire. Furthermore, Likert scales can be applied to many types of questions whereas absolute category rating can only be easily applied to quality evaluations. This is important for non-quality related questions like question 5 of our questionnaire (see Section \ref{questionnaire}).

Thus, although our study is significantly different than other traditional QoE experiments we believe it belongs in the same broad domain and aims at similar goals. Furthermore, our view of QoE is compatible with the Qualinet QoE view. Specifically, our accommodation of long time frames is analogous to their notion of temporal QoE features with so called long term character (such as perceived service reliability over time) \cite{callet2013}. Similarly, the Qualinet view does not specify a specific measurement scale such as MoS. Thus we compare our study to other related QoE studies in section \ref{related_work}.

\subsection{User Retrospective Evaluation}\label{RetroQualityEval}
The retrospective evaluation of a service by a user depends on more than the simple integration of the negative and positive experiences over a service episode. Specifically, temporal effects such as the recency of experiences relative to the entire episode length also affect the evaluation \cite{weiss2014}. A concrete example is the peak-end phenomena that suggests that a significant fraction of an evaluation depends on the peak experience (negative experience with the highest magnitude) and the end experience (last experience of episode) \cite{kahneman1999}.

However, even these well known temporal effects are less applicable if the evaluation occurs significantly after the episode. Specifically, evaluations occurring months or years after the episode rely less on  mental snapshots of experiences and more on general semantic memory \cite{geng2013}. Furthermore, over long periods user behavior adapts to the given system thus producing so called adaptation effects \cite{weiss2014}.

In context of the current study, the situation is further complicated because the user is not evaluating a one-time service episode after a predefined time period but instead a continuous service that the user is utilizing and measuring (in our case of the Netradar app) periodically. In other words, the user must aggregate these long term usage and measurement episodes into an evaluation. Furthermore, some of these episodes are recent (like the measurement directly before the questionnaire) and some typically some take place many months or even a year beforehand.

\section{Data Collection}
This section describes the questionnaire design and deployment along with the mapping of Netradar data and respondent filtering. 
\subsection{Questionnaire Design}\label{questionnaire}
In order to study the aforementioned user satisfaction we implemented a pop-up questionnaire in the Android, iOS, and Windows Phone versions of the Netradar application. Importantly the questionnaire first prompts the user whether they want to participate and thus the questionnaire is completely optional and not required to utilize the application. The participation prompt always appears after the first user-initiated cellular network measurement, and if the user dismisses the prompt then the prompt reappears with a 15\% probability after subsequent user-initiated cellular measurements. If the user completes the questionnaire then the prompt does not reappear for three months. We released the updated Netradar applications that include the questionnaire during the course of summer 2015.

\begin{figure}[!t]
\centering
\includegraphics[width=2.5in]{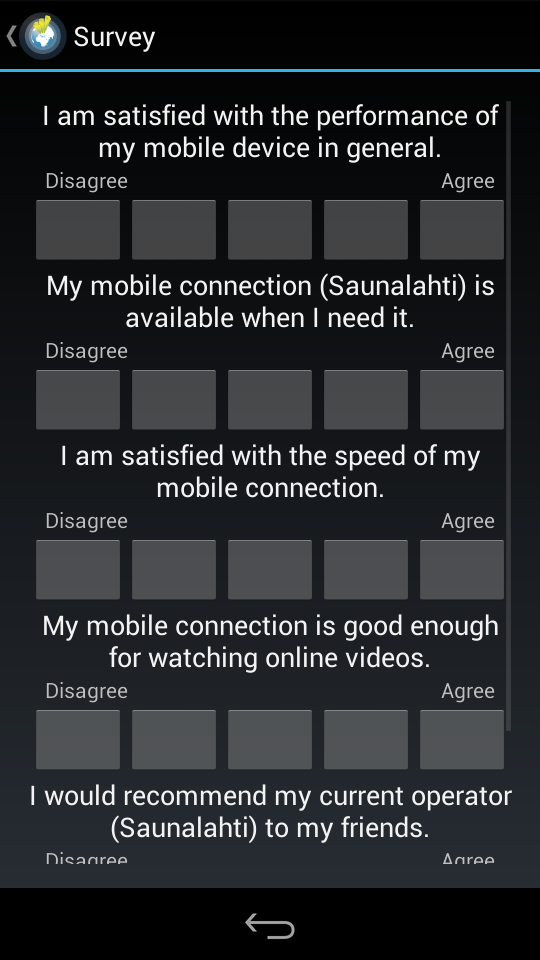}
\caption{Screenshot of the Questionnaire from the Android Netradar App}
\label{screenshot}
\end{figure}

The questionnaire includes five statements with each statement having a five point Likert scale that gauges whether the user agrees or disagrees with the statement. The questions are enumerated below and a screenshot of the questionnaire within the Netradar app is shown in Figure \ref{screenshot}.

\begin{enumerate}[label=Q\arabic*.,itemsep=0pt,labelindent=\parindent, leftmargin=*]
\label{questions}
  \item I am satisfied with the performance of my mobile device in general.
  \item My mobile connection (current operator) is available when I need it.
  \item I am satisfied with the speed of my mobile connection.
  \item My mobile connection is good enough for watching online videos.
  \item I would recommend my current operator (current operator) to my friends.
\end{enumerate}

In terms of question construction, questions Q2 and Q3 were constructed to capture the subjective user satisfaction of network availability (Q2) and network speed (Q3). Question Q4 was constructed to prompt the user to recall specific episodic events (watching videos) in an evaluation rather than potentially relying on the reported network values from the Netradar measurement(s). Finally, Q5 was constructed to assess the relationship between the studied network features and the more tangible business related question of customer recommendation.

We include basic analysis of Q1, Q4, Q5 in this article but leave more extensive analysis of these questions for future work. Thus, we focus primarily on Q2 and Q3 relating to network speed and availability.

Overall as of December 2015, we find that 5,550 users completed the questionnaire. In terms of response rate, the per user response rate is approximately 15.23\%. In other words, 15.23\% of users that saw the questionnaire prompt at least once eventually completed the questionnaire. While the Cronbach's alpha for the four network related questions (Q2-Q5) is 0.8973 suggesting good internal consistency.

\subsection{Data Mapping and Filtering}\label{data_mapping_filtering}
We map each questionnaire respondent to their previous Netradar measurements (abbreviated as meas). We only consider measurements initiated at most 1 year before the respondent completed the questionnaire and only measurements performed on the same network as detected during the questionnaire. The time cutoff decreases the likelihood of, for example, including measurements occurring under different mobile data plans.\footnote{We also performed the analysis with a 2 year cutoff and without any cutoff (thus potentially including measurements since the launch of Netradar in 2013) and did not find large differences.} Furthermore, our unique user identifier is tied to the device itself so all measurements from a user are necessarily from the same device.

The respondents are then filtered to create a more homogeneous subset of respondents for analysis.

First we filter respondents to keep only those based in Finland. We considered a respondent to be Finland-based if the respondent's operator is one of the three major Finnish operators\footnote{The three major operators have over 99\% combined market share of mobile subscriptions in Finland.} (Elisa, Sonera, or DNA). Over 95\% of all respondents as of December 2015 were Finland-based respondents giving a total of 5,274 respondents. We term this resulting dataset as the \textit{complete dataset}.

Second, in order to have an adequate number of measurements per user from which to derive features we further filter the complete dataset by keeping only respondents with at least five measurements and with at least one of those measurements being valid\footnote{We note that for this filtered dataset the median fraction of a user's measurements that are invalid is only 2\% and the maximum fraction is 28\%. Thus no user's measurements is dominated by invalid measurements.} (refer to the definition of a valid measurement in Section \ref{NetradarDesc}) We find that 42.17\% of respondents in the complete dataset meet this criteria thus giving a total of 2,224 respondents. We term the resulting dataset as the \textit{filtered dataset} and we utilize this dataset in all further analysis unless noted otherwise.

In order to illustrate the effect of the aforementioned filtering we perform a basic descriptive comparison between the complete dataset, filtered dataset, and the complement of the filtered dataset\footnote{This is the complement of the filtered dataset with respect to the (superset) complete dataset.} (denoted -filtered) datasets in section \ref{dataset_comparison}.

\section{Network Quality Features}\label{features}
For each respondent we extract a variety of features from their network measurements including the mean, median, and standard deviation of the TCP download, TCP upload, and RTT of their measurements. Additionally we account for the potential temporal effects of retrospective evaluation (refer to Section \ref{RetroQualityEval}) by extracting network features from the most recent valid measurement before the questionnaire (aka last measurement) and the minimum and maximum network values observed over all of the user's valid user-initiated measurements. We note that the last valid measurement is in most cases the user-initiated measurement that prompts the questionnaire, thus the elapsed time between the last measurement and finishing the questionnaire is typically only a few minutes (in 95\% of cases less than 10 minutes, and in 58\% of cases less than 1 minute).

Finally, we also extract a variety of non-network features including the total number of measurements, number of invalid measurements, mean physical speed during measurements, and number of distinct frequently measured locations. These non-network features primarily help differentiate between different types of Netradar users. These different types of users might be satisfied with different levels of network quality and thus accounting for these differences is important.

More specifically, by different types of Netradar users we simply mean that diverse users utilize Netradar for different reasons in a multitude of contexts, thus Netradar users can be differentiated based on these characteristics. For example, from previous experience we postulate that some users utilize Netradar to compare network speeds at their frequent locations (such as work, home, etc.) while other users utilize Netradar to diagnose localized connection or speed problems. Thus we can characterize users based on their primary reason for utilizing Netradar.

For completeness, Table \ref{describeFeaturesTable} lists all the extracted features and the subset of a user's measurements from which each feature is calculated. Next, we briefly describe the methodologies of Netradar for measuring several network and non-network features.
\subsection{Network Features}
\subsubsection{TCP download and upload}
For both download and upload test the Netradar client performs a 10 second data transmission over TCP while sampling the goodput every 50 ms. The 10 second transmission allows the network to reach full capacity even with TCP slow start and long RTTs. The TCP download or upload of the measurement is then calculated as the mean of the goodput samples from the last 5 seconds of the transmission (to avoid including samples from TCP slow start).

\subsubsection{Round Trip Time}
To measure RTT the Netradar client sends a UDP packet to the Netradar server and the server immediately sends a response back to the client. The RTT of the measurement is calculated as the mean RTT of 20 such UDP packet samples. If less than 70\% of packets receive a response then the RTT test of the measurement is considered to have failed. The RTT test is performed after the TCP tests so that the tests do not influence the RTT test.

\subsection{Non-Network Features}
\subsubsection{Invalid Measurement Counts}
As mentioned in section \ref{NetradarDesc}, we also track invalid measurements and the reason for these measurement failures. However, in calculating the counts of invalid measurements we include all invalid measurements that failed for any reason as the Netradar client does not always indicate the reason for a measurement failure.

The invalid measurement counts are included as features because these invalid measurements might act as a proxy for poor network conditions. This is important because due to technical limitations on the iOS platform we do not include signal strength as a feature.

\subsubsection{User-Initiated Measurement Mean Physical Device Speed}\label{physical_device_speed}
At the initiation of the measurement process, the Netradar client requests an update of the current device location including the current physical device speed. Since this location estimate is refined over the length of the measurement\footnote{For example, the accuracy will increase as more fixes are acquired on GPS or GLONASS satellites}, we consider the speed of a single measurement as the speed of the last reported valid location for that measurement. 

All three major platforms provide device speed as part of their Location APIs; however, this speed is provided by the GPS chipset. In other words, in measurements where location is only derived from Wifi or cell-id then device speed is not available.

In light of this, we define a user's mean device speed as the mean speed over all of a user's user-initiated measurements that provide a valid device speed. In other words, a user must have at least one user-initiated measurement with a valid device speed to have a mean device speed. We find that 95.9\% of respondents (in the filtered dataset) meet this requirement and thus have a mean device speed.

Mean device speed is included as a feature because we hypothesize that users that consistently measure at high speeds (where network handoffs are common) might perceive poor network availability (Q2) compared to users with primarily stationary measurements.

\subsubsection{Number of Frequently Measured Locations}
For each user we approximate the number of frequently measured locations as the number of distinct spatial clusters of their user-initiated measurements.

To find the number of clusters we use DBSCAN clustering \cite{ester1996} with the orthodromic (great circle) distance as the distance metric. DBSCAN clustering is a density based clustering method that has the advantage that the number of clusters does not need to be specified beforehand (in contrast to K-Means). In terms of utilized parameters, the minimum number of measurements for a dense region is 3 and the $\varepsilon$-neighborhood is 500 meters. The location of any single measurement is considered the reported location with the best accuracy estimation.

The user must have at least one valid user-initiated measurement to be able to cluster since invalid measurements (typically) do not provide location data. We find that 96.4\% of respondents (in the filtered dataset) meet this requirement and thus have a valid number of frequently measured locations.

These measurement locations likely represent meaningful places for users such as home or work. Thus the rationale for inclusion is that this feature might be a proxy for different types of Netradar users as explained in the example from Section \ref{features}.

\subsubsection{User-Initiated Measurement Burstiness}
For each user we calculate the temporal burstiness, $B$ \cite{goh2008}, of the user's user-initiated measurements. $B$ is defined in terms of the sample mean, $m_{\tau}$, and sample standard deviation, $\sigma_{\tau}$, of the interevent time distribution $\tau$ as detailed in Equation \ref{eq:burstiness}. $B$ can vary between -1.0 and 1.0, with a value of 1.0 indicating very bursty, 0.0 indicating random (Poissonian), and -1.0 indicating periodic. 

The user must have at least three user-initiated measurements to calculate a temporal burstiness measure because the sample standard deviation of a one sample interevent distribution is undefined. We find that 99.1\% of respondents (in the filtered dataset) meet this requirement and thus have a temporal burstiness measure.

\begin{equation}\label{eq:burstiness}
B \equiv \frac{(\sigma_{\tau} - m_{\tau})}{(\sigma_{\tau} + m_{\tau})}
\end{equation}

Temporal burstiness is included as a feature because we hypothesize that frequent bursts of user-initiated measurements might indicate intervals of network problems or performance degradation whereas periodic measurements might indicate simple occasional checks of network performance. Thus, similarly to the number of frequently measured locations, we hypothesize that burstiness might be a proxy for different types of Netradar users.

\subsubsection{Mobile Device Type and Category}\label{device_type_category_desc}
Finally, the device model should be included as a feature because different device properties such as device quality or device type (smartphone or tablet) might have an effect on satisfaction (even network satisfaction). However, we cannot include mobile device model directly as a categorical variable because of the large number of device models with very few users. Thus we create and include two mobile device based features to account for differences in device type and device quality. 

First we collect publicly available ancillary data about each device model that appears in the filtered dataset including device type\footnote{We define a device as a tablet if screen size in diagonal inches $\ge$ 7.0 and otherwise a smartphone.} (smartphone or tablet), screen size in diagonal inches, screen pixel density in pixels per inch, number of CPU cores and frequency of each core in GHz, and finally theoretical floating point operations of GPU in GFLOPs.

Then for smartphone and tablet devices separately we cluster the devices into three groups based on four standardized features (screen size, screen pixel density, total frequency of CPU (cores times frequency per core), and GPU GFLOPs) with k-means clustering (500 replicates with final clustering from the replicate with lowest within-cluster sums of point-to-centroid distances). The three groups are termed high, medium, and low to represent device quality relative to the highest specified devices today. We include this custom device quality categorical variable as one of the device based features.

The other included feature is a binary device type feature that is one if the device is a tablet and zero if the device is a smartphone.
\begin{table}[!t]
\centering
\begin{threeparttable}
\caption{User Features and the Subset of User's Measurements from which Each Feature is Calculated or Derived}
\label{describeFeaturesTable}
\begin{tabular}{@{}ll@{}}
\toprule
\textbf{Feature} & \textbf{Calculated From}\\ 
\midrule
Mean TCP Down & Valid Measurements\\
Median TCP Down & Valid Measurements\\
Standard Deviation TCP Down & Valid Measurements\\
Minimum TCP Down & User-Initiated Valid Measurements\\
Maximum TCP Down & User-Initiated Valid Measurements\\
Last TCP Down & Last Valid Measurement\tnote{a}\\
Mean TCP Up\tnote{b} & Valid Measurements\\
Median TCP Up\tnote{b} & Valid Measurements\\
Standard Deviation TCP Up\tnote{b} & Valid Measurements\\
Minimum TCP Up & User-Initiated Valid Measurements\\
Maximum TCP Up & User-Initiated Valid Measurements\\
Last TCP Up & Last Valid Measurement\tnote{a}\\
Mean RTT & Valid Measurements\\
Median RTT & Valid Measurements\\
Standard Deviation RTT & Valid Measurements\\
Minimum RTT & User-Initiated Valid Measurements\\
Maximum RTT & User-Initiated Valid Measurements\\
Last RTT & Last Valid Measurement\tnote{a}\\
\midrule
Device Platform\tnote{c} & Any Measurement\\
Device Type\tnote{d} & Any Measurement\\
Device Quality\tnote{d} & Any Measurement\\
\midrule
Temporal Burstiness & User-Initiated Measurements\\
Number Frequently Measured Locations & User-Initiated Valid Measurements\\
Mean Physical Device Speed & User-Initiated Valid Measurements w/ Valid Device Speed \\
Number User-Initiated Measurements & User-Initiated Measurements\\
Number Measurements & All Measurements\\
Number Invalid Measurements & Invalid Measurements\\
Number Invalid User-Initiated Measurements & User-Initiated Invalid Measurements\\
\bottomrule
\end{tabular}
    \begin{tablenotes}
      \small
      \item[a] As previously mentioned, the last valid measurement is almost always the user-initiated measurement that prompts the questionnaire and thus is also user-initiated.
      \item[b] In cases of Windows Phone devices, the mean, median, and standard deviation TCP upload features are based on only user-initiated valid measurements. This is due to a limitation in the Windows Phone platform such that background tasks can only run continuously for a maximum of 25 seconds, thus the Netradar Windows Phone client only measures TCP down and RTT in background measurements.
      \item[c] Android, iOS, or Windows Phone
      \item[d] These are custom features derived from the device model and ancillary data. Refer to Section \ref{device_type_category_desc}.
    \end{tablenotes}
\end{threeparttable}
\end{table}

\section{Analysis}\label{analysis}
This section describes the analysis of the data including the effect of the previous respondent filtering, basic statistics of the filtered dataset, and the creation of ordinal regression models for each question.
\subsection{Basic Descriptive Statistics}
\subsubsection{Dataset Comparison}\label{dataset_comparison}
Before exploring the filtered dataset in depth, we compare the complete dataset with both the filtered dataset and its complement to check for obvious biases that the filtering might cause.  Table \ref{dataset_comparison_table} details several descriptive statistics for all three datasets (complete, filtered and -filtered (complement of filtered)). We find no clear biases in terms of, for example, platform, operator, or median TCP down.

The bottom two rows of Table \ref{dataset_comparison_table} illustrate examples of two reasons for the filtering.  First, dispersion features such as standard deviation are less useful when the number of measurements is small.  Second, features that describe different aspects of the same distribution, such as central tendency (median) and last observation (last) of TCP down, are highly correlated and again less useful (in subsequent models) when the number of measurements is small.

\begin{table}[!t]
\centering
\begin{threeparttable}
\caption{Descriptive Statistics of Several Features for Different Datasets}
\label{dataset_comparison_table} 
\centering
\begin{tabular}{@{}llllll@{}}
\toprule
\textbf{Statistic} & \textbf{Complete} & \textbf{Filtered} & \textbf{-Filtered\tnote{a}} \\
 \midrule
Number Respondents & 5274 & 2224 & 3050\\ 
Total Number Measurements & 295065 & 290142 & 4923\\
Android (\%) & 43.84 & 42.27 & 44.98\\ 
Windows Phone (\%) & 31.15 & 32.64 & 30.07\\
iOS (\%) & 25.01 & 25.09 & 24.95\\ 
Operator 1 (\%) & 45.81 & 45.82 & 45.80\\ 
Operator 2 (\%) & 34.32 & 35.03 & 33.80\\ 
Operator 3 (\%) & 19.87 & 19.15 & 20.39\\ 
Median of Median TCP Down (Mbps) & 10.66 & 10.40 & 10.92\\ 
\midrule
Median of Std. Dev. TCP Down (Mbps) & 1.55 & 7.70 & 0.00\\ 
Correlation\tnote{b}\ \ (Median and Last TCP Down) & 0.72 & 0.51 & 0.89\\ \bottomrule
  \end{tabular}
  \begin{tablenotes}
      \small
      \item[a] The complement of the filtered dataset within the complete dataset
      \item[b] Kendall Tau-b Correlation
\end{tablenotes}
\end{threeparttable}
\end{table}

\subsubsection{Filtered Dataset}
The filtered dataset consists of 2,224 distinct respondents and 290142 corresponding measurements. We find that 66,222 (about 23\%) of these measurements are user-initiated measurements and 26,480 (about 9.5\%) are invalid measurements. While the respondents devices represent about 200 distinct models from 24 different vendors. We find 13\% of respondents have tablet devices and 87\% smartphones. The distributions between the device quality variable for respondents smartphones are high: 49.43\%, medium: 41.16\%, low: 9.41\% and for tablets are high: 41.38\%, medium: 41.46\%, low: 15.17\%. Further user level statistics for the numeric features are detailed in Table \ref{descriptive_stats_table}.

\begin{table}[!t]
\caption{User Level Descriptive Statistics of Numeric Features}
\label{descriptive_stats_table} 
\centering
\begin{tabular}{@{}llllll@{}}\hline  
\toprule
\textbf{Feature} & \textbf{Mean} & \textbf{St. Dev.} & \textbf{Median} \\ \midrule
Mean TCP Down (Mbps) & 16.65 & 14.39 & 12.58\\ 
Median TCP Down (Mbps) & 15.38 & 15.39 & 10.40\\ 
St. Dev. TCP Down (Mbps) & 10.09 & 8.48 & 7.70\\ 
Last TCP Down (Mbps) & 17.85 & 20.25 & 10.15\\ 
Min TCP Down (Mbps) & 4.20 & 7.77 & 1.49\\ 
Max TCP Down (Mbps) & 36.07 & 29.82 & 27.77\\ 
Mean TCP Up (Mbps) & 5.87 & 6.25 & 3.37\\ 
Median TCP Up (Mbps) & 4.94 & 6.78 & 2.17\\ 
St. Dev. TCP Up (Mbps) & 4.62 & 4.52 & 3.08\\ 
Last TCP Up (Mbps) & 6.03 & 8.67 & 2.24\\ 
Min TCP Up (Mbps) & 1.00 & 2.70 & 0.26\\ 
Max TCP Up (Mbps) & 15.24 & 14.05 & 9.87\\ 
Mean RTT (ms) & 106.16 & 159.46 & 57.83\\ 
Median RTT (ms) & 61.34 & 101.73 & 40.5\\ 
St. Dev. RTT (ms) & 138.95 & 264.36 & 38.28\\ 
Last RTT (ms) & 106.03 & 323.24 & 41.00\\ 
Min RTT (ms) & 37.26 & 46.99 & 28.00\\ 
Max RTT (ms) & 611.17 & 1087.41 & 154.65\\ 
Temporal Burstiness & 0.26 & 0.19 & 0.26\\ 
Number Frequently Measured Locations & 1.43 & 2.14 & 1.00\\ 
Mean Physical Device Speed (m/s) & 0.75 & 2.25 & 0.04\\ 
Number Measurements & 130.46 & 1564.08 & 15.00\\   
Number User-Initiated Measurements & 29.78 & 61.89 & 13.00\\   
Number Invalid Measurements & 12.47 & 165.89 & 1.00\\   
Number Invalid User-Initiated Measurements & 2.73 & 8.89 & 1.00\\   
\bottomrule
  \end{tabular}
\end{table}

\begin{figure}[!t]
\includegraphics[width=6.3in]{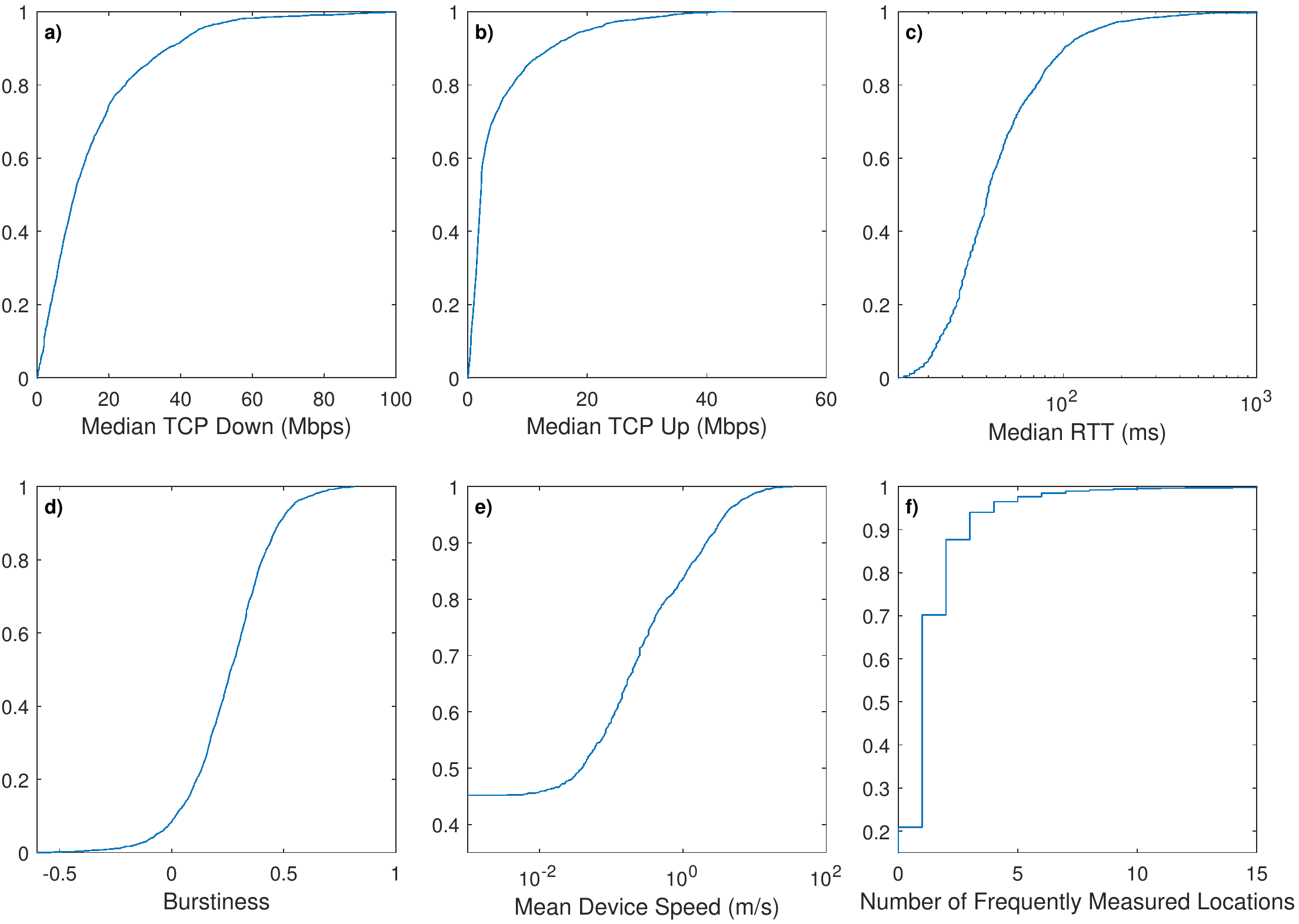}
\caption{Empirical Cumulative Distribution Functions for Several Selected Features: a) Median TCP Down, b) Median TCP Up, c) Median RTT, d) Temporal Burstiness Measure for User Initiated Measurements, e) Mean Device Speed for User Initiated Measurements, f) Number of Frequently Measured Locations}
\label{ecdfs}
\end{figure}

We also detail the full distributions of several interesting features in Figure \ref{ecdfs} to elaborate beyond the given basic statistics. The median TCP down and up distributions indicate, as expected, concentrated distributions that imply that most users have moderate TCP download/upload goodputs and while a small fraction of users have very high goodputs. The RTT distribution also indicates high concentration with a large majority having median RTTs between 20 and 100 ms. The burstiness distribution shows that almost all users have moderate temporal burstiness (0.20 to 0.60) in their user initiated measurements. These burstiness coefficients are similar to coefficients found in other human activities including sending email, initiating phone calls, and printing \cite{goh2008}. In terms of mean physical device speed, the large majority of the users mean device speed (\textgreater 80\%) is less than 1 meter/second (3.6 km/h) indicating primarily stationary measurements. Though we do find 3\% of users have mean device speeds greater than the average human sprinting speed (over short distances) of 6.71 m/s (24.14 km/h), thus suggesting vehicular measurements. The frequent location distribution indicates that a large majority of users (about 87\%) frequently measure two or less distinct locations, whereas only a very small fraction measure more than five distinct locations. 

\subsubsection{Questionnaire Response Cross-Correlations}
We examine the Spearman rank cross correlations between the questionnaire answers for respondents. Table \ref{answers_correlations} details these correlations and all correlations are statistically significant at the 0.1\% level. 

\begin{table}[!t]
\caption{Questionnaire Response Spearman Rank Cross-Correlations}
\label{answers_correlations} 
\centering
\begin{tabular}{@{}llllll@{}}  
\toprule
 & Q1  & Q2  & Q3  & Q4  & Q5  \\ 
 \midrule
Q1 & 1.00 & & & &\\  
Q2 & 0.40 & 1.00 & & & \\  
Q3 & 0.51 & 0.63 & 1.00 & & \\  
Q4 & 0.50 & 0.59 & 0.78 & 1.00 & \\  
Q5 & 0.43 & 0.70 & 0.74 & 0.71 & 1.00 \\  
\bottomrule
  \end{tabular}
\end{table}

Unsurprisingly, we find the largest correlation (0.78) between Q3 related to network speed and Q4 related to watching online videos. Furthermore, we find very high correlations between Q3 and Q5 related to recommending operator to friends and between Q2 related to network availability and Q5. These high correlations between network measures and recommending operator helps justify the importance that operators put on network quality. Though importantly the sample of Netradar users is biased towards users that already have a general interest in network quality.

\subsubsection{Numeric Feature Cross-Correlations}
Next, we examine the cross correlations between the numeric features described in Section \ref{features}. For each pair of features we calculate the Kendall Tau-b correlation and associated statistical significance. The correlations and significance levels are detailed in Table \ref{numeric_feature_correlations} and Table \ref{numeric_feature_significances} respectively in the Appendix. 

Overall, we find that 96.7\% of correlations between network related features are statistically significant at at least the 1\% level and 95.4\% are significant at at least the 0.1\% level. In other words, as expected, network performance measures tend to move in unison.

Interestingly, we find that correlations between median network features and corresponding last network features (i.e. between median TCP down and last TCP down) are positive and significant ($p \leq 0.1\%$) but not at overwhelmingly high levels. These correlations for TCP down, up, and RTT are 0.51, 0.46, and 0.56 respectively. Thus, both median and last features will likely be viable in the same statistical model (in other words, their collinearity will not be too high).

The correlations involving non-network features are also interesting, especially two of the non-network, non-count features: burstiness and number of frequently measured locations. 

In terms of temporal burstiness, we find that burstiness is weakly but significantly ($p \leq 0.1\%$) correlated with several network features. All of the correlation signs indicate higher burstiness is associated with poorer network quality. This supports the hypothesis that frequent bursts of user-initiated measurements might indicate intervals of network problems or performance degradation. 

The number of frequently measured locations and all maximum, minimum, and standard deviation network features are significantly ($p \leq 0.1\%$) correlated. Intuitively, as a user measures more diverse spatially distinct locations the user is more likely to encounter a location with very good or very bad network performance.

\subsection{Ordinal Logistic Regression Models}
Finally, we construct a series of ordinal logistic regression models to determine which features are statistically significant with respect to predicting responses for each questionnaire question. We first briefly explain our procedures for multicollinearity testing, missing data imputation, and model building. Then we present the resulting models and describe the results.

For convenience hereafter we refer to the Likert scale values for the questions by numerical shorthand values of 1 to 5 with 1 representing disagree and 5 representing agree. Also we denote aggregations of scale values by parentheses; in other words, (1,2) vs (3,4,5) indicates a simple dichotimization of the responses.

\subsubsection{Model Specification and Building}
In terms of multicollinearity, highly correlated features in a regression model can cause problems including inflated standard errors. Thus we perform multicollinearity diagnostics on the features using the \texttt{collin} \cite{ender2010} command in Stata 13 which reports variance inflation factors (VIF) for each feature. We utilize a incremental feature elimination approach in which we eliminate the feature with the highest VIF over 10 until none of the remaining features have VIFs over 10. The VIF cutoff of 10 is a widely used guideline originally proposed in \cite{marquardt1970}. Through this procedure we remove all three mean network features (TCP upload, download, and RTT) and maximum TCP download (which was highly correlated with standard deviation of TCP download) from model building.

In terms of missing data, as described in Section \ref{features} three features have missing data for a fraction of users: temporal burstiness, number of frequently measured locations, and mean device speed. We utilize sequential imputation using chained equations with each feature modeled through predictive means matching (due to non-normality of the features) \cite{white2011}. We use the Stata prefixes \texttt{mi impute} for multiple imputation and \texttt{mi estimate} to utilize the imputations in the subsequent model building. 

For building each model, we utilize the user defined \texttt{gologit2} \cite{williams2006} command with the \texttt{autofit} option in Stata 13. The command performs a stepwise procedure that tests each feature for the proportional odds assumption (also known as parallel lines assumption) and subsequently builds the appropriate model type. Through this procedure, we build five partial proportional odds models as in each case some but not all features meet the assumption.

\subsubsection{Model Results}
In terms of overall model results, Table \ref{sig_levels} details the significance levels of each feature for each model.  Furthermore, Tables \ref{Q2_model_coeff} and \ref{Q3_model_coeff} in Appendix A give detailed model coefficients and standard errors for models of Q2 and Q3.

\paragraph{Network Availability}
In terms of availability, the Q2 model suggests that minimum TCP down, number of frequently measured locations, network operator, and device type are significant at at least the 1\% level.

Intuitively, low minimum TCP down values might imply poor network conditions and low availability in at least one location. Furthermore, given that users typically measure only a few significant locations, if these poor conditions are prevalent at a significant location then poor availability might be frequently noticed. 

In terms of number of frequently measured locations, we find that the higher the number of locations the higher the predicted Likert score (see coefficient in Table \ref{Q2_model_coeff}). This supports our postulation that this feature is a rough proxy for the different types of Netradar users. More specifically, we hypothesize that some users measure in multiple locations to simply compare speeds, whereas other users measure in only a single location often to troubleshoot specific network problems.

The significance of mobile network operator despite the inclusion of many network features might be due to differences in coverage between networks since we did not include signal strength directly as a feature (and the number of invalid measurements is only a rough proxy\footnote{In fact, the invalid measurement features are not significant even at a 10\% level in terms of availability (Q2)}). We look to include signal strength in future work if platform specific constraints allow. 

Interestingly, though mean physical device speed is not significant in our baseline (1,2) vs (3,4,5) comparison, we find it is significant ($p \leq 1\%$) in the (1) vs (2,3,4,5) comparison (see Table \ref{Q2_model_coeff}) suggesting that mean physical device speed is predictive in the case of predicting very poor availability. This supports the postulation from Section \ref{physical_device_speed}.

Finally, the significance of the device type feature is likely due to device mobility since tablets are primarily utilized at home where a user is likely to have chosen the best available network whereas smartphones are utilized in diverse locations with more diverse network conditions.

\paragraph{Network Speed}
In terms of network speed, the Q3 model indicates that minimum, median, and last TCP down features are significant at at least the 1\% level. Thus these results suggest that both the peak-end effect and integration over measurements play a role in determining user satisfaction with network speed. The observation of both these effects indicates that the evaluation of network speed satisfaction is relatively complex.

\paragraph{General Observations}
Overall, we find that the minimum, median, and last TCP down features are significant across many models. In fact, the high significance ($p \leq 0.1\%$) of last TCP down in so many models especially highlights the importance of temporal effects in such satisfaction evaluations.

Interestingly, we find that very few of the TCP upload features are significant in any of the models. This result can be understood in context that for a typical user the download share of their mobile traffic is much larger than the upload share. For example, Sandvine \cite{sandvine2014} reports that for a median European user the download-upload mobile traffic split is 88\%-12\%.

Furthermore, relatively few of the latency based features are highly significant. We postulate that many users might not be aware of the nature of latency and high latency might not directly effect their opinions. In addition, latency below a certain threshold is valuable in only a relatively small subset of applications, thus in our case the effect of latency might not be perceivable to many users. In a specific example, Wac et al. \cite{wac2015} describe a variety of latency requirements for different mobile applications from literature with the lowest requirement level being 100 ms for first person shooter video games. Our median RTT distribution (Figure \ref{ecdfs}c) illustrates that about 90\% of our respondents already have median RTT below 100 ms.

In terms of non-network features, the number of user-initiated invalid measurements is only weakly significant in a few models. The weak significance of invalid measurement counts might be partly due to the relative infrequency of invalid measurements overall. Specifically, 46\% of users do not have any invalid measurements at all and for the median user with 13 user initiated measurements only 1 of these are invalid.

Unsurprisingly, the custom device quality feature is highly significant in the Q1 model as would be expected. However, interestingly the feature is not significant in any of the other models, thus suggesting that device quality is, in and of itself, not predictive of network availability or speed satisfaction.

Finally, the adjusted McFaddens pseudo $R^{2}$ of the partial proportional odds models are 0.050, 0.035, 0.052, 0.079, and 0.048 for Q1-Q5 respectively\footnote{If we collapse the ordinal dependent variable to a binary dependent variable by considering only the baseline (1,2) vs (3,4,5) comparison we find McFaddens pseudo $R^{2}$ values of 0.075, 0.067, 0.100, 0.139, and 0.083 for Q1-Q5 respectively}. These goodness of fit values indicate that the models provide only a basic fit and that the models cannot explain all of the variance. In other words, other factors undoubtedly play a role in these evaluation and thus conclusions based on the model significances must be seen as only a part of the puzzle. Also we note that McFadden pseudo $R^{2}$ values are not directly comparable to ordinary regression $R^{2}$ values as pseudo $R^{2}$ values are typically much lower. For reference our pseudo $R^{2}$ values of 0.067 and 0.100 for the binary Q2 and Q3 models roughly translate to ordinary $R^{2}$ values of 0.10 and 0.22 respectively  \cite{mcfadden1975}. 

\begin{table}[!t]
\centering
\begin{threeparttable}
\caption{Feature Significance Levels\tnote{c} for Logit Models}
\label{sig_levels} 
\begin{tabular}{@{}llllll@{}} 
\toprule
\textbf{Feature} & \textbf{Q1} & \textbf{Q2} & \textbf{Q3} & \textbf{Q4} & \textbf{Q5} \\ \midrule
Median TCP Down & & * & ** & ** & \\ 
St. Dev. TCP Down & * & & & & *\\ 
Last TCP Down & ***\tnote{a} & & *** & ***\tnote{a} & ***\tnote{a}\\ 
Min TCP Down & & **\tnote{a} & ***\tnote{a} & * & \\ 
Median TCP Up & ** & & & * & \\ 
St. Dev. TCP Up & & & & & \\ 
Last TCP Up & & & & & \\ 
Min TCP Up & & & & & \\ 
Max TCP Up & & & & & \\ 
Median RTT & & & & & \\ 
St. Dev. RTT & & & & & \\ 
Last RTT & * & * & & * & \\ 
Min RTT & & & & & *\tnote{a}\\ 
Max RTT & & * & & *\tnote{a} & *\\ 
Temporal Burstiness & & & & ** & *\\ 
Number Frequently Measured Locations & & ** & * & ** & **\\ 
Mean Device Speed & & & & & \\
Number Measurements & & & & & \\ 
Number User-Initiated Measurements & * & & & & \\ 
Number Invalid Measurements & & & & & \\
Number Invalid User-Initiated Measurements & & & * & * & *\\ 
Device Platform & *\tnote{b} & * & * & *\tnote{b} & ***\tnote{b}\\ 
Mobile Network Operator & & ** & & * & ***\tnote{b}\\ 
Device Quality & *** & & & & \\ 
Device Type (Tablet or Smartphone) & & ** & & & \\ 
\bottomrule
  \end{tabular}
    \begin{tablenotes}
      \small
      \item[a] Feature did not satisfy parallel lines assumption, reported significance level is for baseline of (1,2) vs (3,4,5). See Appendix for full models of Q2 and Q3.
      \item[b] Feature did not satisfy parallel lines assumption and feature is categorical, reported significance level is for baseline of (1,2) vs (3,4,5) for most significant comparison between any two categories. See Appendix for full models of Q2 and Q3.
      \item[c] $\ast$ : 5\%, $\ast\ast$ : 1\%, $\ast{\ast}\ast$ : 0.1\%
    \end{tablenotes}
\end{threeparttable}
\end{table}

\subsubsection{Predicted Probabilities as Effect Size}
In order to convey the magnitude of the effects of significant features we illustrate the model predicted probabilities of the dependent variables while varying a single significant feature and keeping all other features at observed values. We utilize the user written \texttt{mcp} \cite{royston2013} command in Stata 13 to perform the analysis.

\begin{figure}[!t]
\includegraphics[width=6.3in]{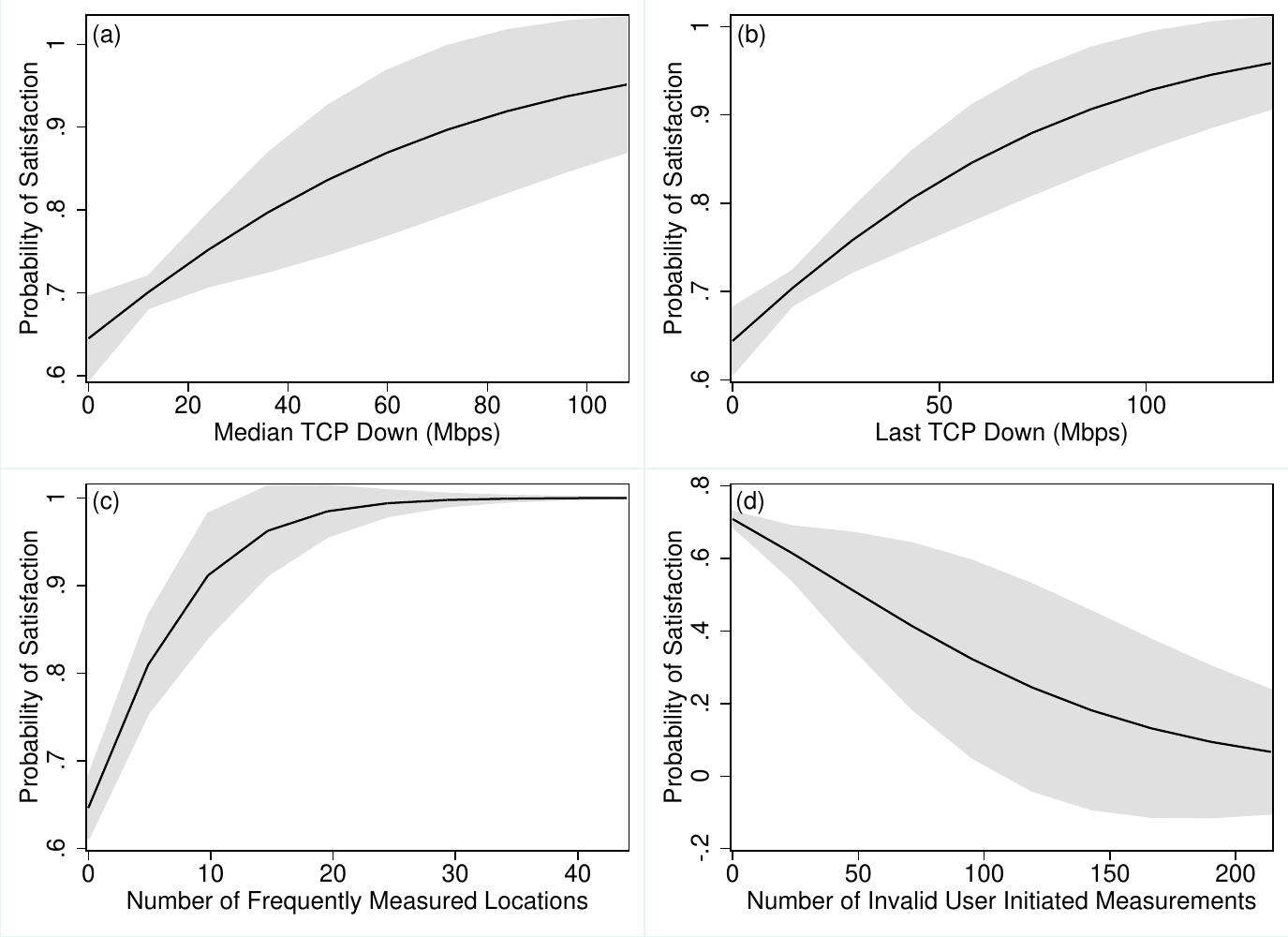}
\caption{Predicted Probability Functions for Several Significant Features of Q3: (a) Median TCP Down, (b) Last TCP Down, (c) Number of Frequently Measured Locations, (d) Number of Invalid User Initiated Measurements. The Stata command \texttt{mcp} was utilized to compute the probability functions \cite{royston2013}}
\label{predicted_probs_q3}
\end{figure}

Importantly, since adjusted probabilities in ordinal cases are only intuitive for binary outcomes we again collapse the ordinal dependent variable to a binary dependent variable by considering only the (1,2) vs (3,4,5) comparison. We term the (3,4,5) state as satisfaction. We then re-estimate the models as simple binary logistic regression models. Furthermore, since all models are simple binary logistic regression models we no longer need to differentiate between normal and generalized models. In addition due to technical limitations we do not utilize multiple imputation but instead utilize complete case analysis. This change though does not substantially alter the outcomes.

We focus on network speed satisfaction (Q3), Figure \ref{predicted_probs_q3} illustrates the predicted probabilities of satisfaction for Q3 over four features: median TCP down, last TCP down, number of frequently measured locations, and number of invalid user initiated measurements. As expected in both the median and last TCP down cases we find that the probability of satisfaction increases rapidly as the goodput increases but shows constant diminishing returns. In terms of locations, we find that probability of satisfaction increases strongly with the number of frequently measured locations. Finally, the number of user initiated invalid measurements suggests a strong effect but as previously discussed relatively few users have significant numbers of invalid measurements and thus no strong conclusions can be drawn.

\subsection{Representativeness of Respondents}
The sample of Netradar users is undoubtedly not representative of the Finnish smartphone market at large because Netradar primarily attracts advanced users that are interested in measuring their mobile connection or troubleshooting network connection problems. However, the degree of representativeness can be approximated by utilizing supplementary data available for the Finnish mobile market.  Specifically, we can use publicly available data on the distribution of top mobile phone models in use on major Finnish mobile networks from a recent report \cite{vesselkov2016}. The report data was collected in September 2015 and thus relatively closely matches the questionnaire deployment.

We take the top 13 smartphone devices from both the filtered dataset and the Finnish mobile network report. We then calculate the average screen size, screen pixel density, total frequency of CPU (cores times frequency per core), and GPU GFLOPs for all devices on the lists. Table \ref{netradar_device_repre} details the differences between the mean values of these specifications.  We find that the Netradar respondents have higher device specifications in all areas thus supporting that these are likely advanced users. In fact, the differences suggest that Netradar users are at least one mobile device generation ahead of the average Finnish smartphone user overall.

\begin{table}[!t]
\caption{Average Smartphone Specs for Top 13 Device Models of Netradar Respondents and all Finnish Mobile Network Users}
\label{netradar_device_repre}
\centering
\begin{tabular}{@{}llllll@{}}  
\toprule
 & Screen Size (in.) & Pixel Density (ppi) & Total Freq (GHz) & GPU (GFLOPs)\\
 \midrule
Netradar Respondents & 4.70 & 346.00 & 4.80 & 71.61\\  
Finnish Network Users & 4.19 & 280.85 & 2.92 & 34.74\\ 
\bottomrule
  \end{tabular}
\end{table}

\section{Discussion}\label{discussion}
We briefly discuss an issue related to respondents interpretation of the questionnaire questions that helps frame the overall results.

Specifically, we have to consider the interpretation of the questions by the respondents given the Netradar and non-Netradar information they might consider in their evaluations. As an example, a respondent might interpret Q3 as: ``are you satisfied with the network speed given the Netradar measurements?''. In this case the respondent might consider the speed provided by the Netradar measurements and compare against the marketed speed for their data plan and then give an appropriate rating. Another respondent might interpret Q3 as: ``are you satisfied with the network speed given the mix of applications you utilize such that these applications perform adequately?''. In this case the respondent might consider network performance during specific network usage episodes. Furthermore, another respondent might consider a combination of Netradar measurements and network usage episodes. In other words, understanding the information the respondent is utilizing in evaluating the statement can helps with the interpretation of the results.

Generally, given the diversity in the relationship between download goodput and Q3 answers, we postulate that many users consider some combination of Netradar measurements and network usage episodes. However further research is required to understand and disentangle these factors. Additionally, in terms of studying the sub-population of mobile network users that measure the network and observe these measurements, this interpretation ambiguity is unavoidable.
\section{Related Work}\label{related_work}
The related work can be divided into several different areas.
\subsection{Device based Mobile Network Measurements}

A large number of works have developed or utilized similar device based mobile measurement applications \cite{goel2015}. These applications vary slightly in the detailed measurement methodology (again refer to Table 2 in \cite{goel2015}), however, their results are still roughly comparable. The applications most similar to Netradar include Open Signal, Root Metrics, and Sensorly. However, these applications typically do not survey users as their focus has tended to be network-centric.

For example Nikravesh et al. \cite{faloutsos2014} studied mobile network performance in three major US cities through two device based measurement applications. They found several operator specific network impairments. In terms of differences, their network measurements were initiated at app defined intervals, whereas our measurements are a combination of user initiated and app initiated (given user defined parameters). And again they do not survey users as their focus is network-centric.

\subsection{Device based Mobile Network Measurements with QoE}
A few studies have combined device measurement applications and QoE surveys. Kobayashi et al. \cite{kobayashi2015} studied the mobile network performance during certain applications and the user reported QoE degradation during those applications. They found that users reported QoE degradation depending on the usage context even given similar network conditions. In comparison to our approach, they collect passive measurements whereas our application performs active measurements, and they ask users about QoE on a per app basis whereas we ask for aggregate network speed and availability satisfaction. Furthermore, they utilize a relatively small panel of 44 users, whereas our final filtered dataset contains 2,224 users.

Similarly \cite{casas2015b} also studied mobile network bandwidth estimations during certain applications and per app user reported QoE evaluations. They report field results that roughly correspond to lab results they previously found in \cite{casas2015a} and that are described in the next section.

\subsection{Laboratory based Mobile Network Scenarios with QoE}
Several studies have examined related topics in laboratory settings. A few studies are highlighted.

Casas et al. \cite{casas2015a} studied the download bandwidth required for different QoE levels for a variety of popular smartphone applications in a lab experiment. They found that a bandwidth of 4 Mbps was sufficient to reach MoS saturation (and thus near best achievable quality) for all of the studied popular apps. A subtle but important difference though between \cite{casas2015a} and our approach is that our users knew the objective network speed (since they were actively measuring) and thus were likely to take this speed into consideration whereas the lab subjects in \cite{casas2015a} were unaware of actual speed.

Similarly Hosek et al. \cite{hosek2014} explicitly modeled the mean opinion score as a function of download bandwidth and delay for three common mobile Internet scenarios (web browsing, file download, and file upload). The model was again based on similar lab testing.

Also, Agboma and Liotta \cite{agboma2010} analyzed and modeled the user perceived acceptability for a variety of different types of video content (sports, movies, cartoons, etc.) in different device contexts (laptop, phone, PDA) under different simulated network conditions (specifically different encoding bit and frame rates). They found significant differences in the requirements for acceptability depending on the content and device type. Furthermore, related work argues that these lab based results can be applied more broadly through application of automated machine learning to the lab data \cite{menkovski2010}.

\subsection{Combined Laboratory based and Device based Mobile Network Scenarios with QoE}

Schatz and Egger \cite{schatz2011} studied both laboratory and device based QoE trials where users reported MOS and acceptability scores for different download speeds in web browsing and file downloading contexts on PCs. The download speeds were shaped by custom software on the user's PCs. The found that that for web browsing the lab and device results agreed but for file downloading the results significantly differed likely because users in the field were able to distract themselves during the downloading time.

\subsection{Questionnaire based User Satisfaction of Mobile Services}
Finally, several studies have examined user satisfaction with mobile services through consumer questionnaires \cite{ozer2013,kuo2009,turel2006}. For example, {\"O}zer et al. \cite{ozer2013} utilized exploratory factor analysis of questionnaire data to determine the factors affecting customer satisfaction with mobile service. They found that service availability significantly affects customer satisfaction, though their definition of service availability encompasses both our concepts of network speed and availability. However, none of these studies use network measurements to evaluate network quality. Furthermore, these studies stem primarily from the business and marketing domains rather that the technical domain as the other related studies. 

\subsection{General QoS to QoE Prediction}
QoS is a well studied concept that focuses on system oriented and objectively measurable metrics such as network goodput and latency; whereas QoE is a newer concept that focuses on subjective end-user metrics and accounts for contextual factors \cite{varela2014}. In light of these concepts, our result models can be viewed as related to QoS to QoE prediction models. Thus we briefly describe the applications and issues of QoS to QoE prediction.

In terms of applications, QoS to QoE prediction can help network and application management systems efficiently allocate resources to maximize QoE \cite{schatz2014}. Specifically the QoS to QoE prediction models allow these systems to take into account the non-linear relationships between QoS metrics and user QoE (these non-linearities often result from properties of human perception systems) and known contextual information (such as location and time) \cite{varela2014}. For example, beyond certain limits or in certain contexts increases or decreases in certain QoS metrics will not effect user perceived quality and the systems can take these limits into account. Though a limitation of such systems is that varying applications in varying contexts have varying QoS to QoE relationships and additionally many context factors are difficult to measure (for example task and social context as discussed in \cite{reiter2014}). Thus apriori prediction of QoS to QoE in general is difficult.

In terms of further references, Schatz et al. \cite{schatz2013} and Varela et al. \cite{varela2014} survey and provide discussion about the relationship between QoS and QoE. While Alreshoodi and Woods \cite{alreshoodi2013} review mathematical models for mapping network QoS metrics to QoE. 

\section{Conclusion}\label{conclusions}

In this article, we studied the significant predictors of user satisfaction with regard to several mobile network concepts in the context of end user mobile network measurements. In other words, we studied these predictors for users that themselves actively measure the network and observe these measurement results. The analyzed dataset combines user questionnaire responses and the corresponding questionnaire respondents network measurements. In terms of size, the dataset consists of 2,224 Finland-based respondents with 290,142 network measurements performed over 12 months.

In terms of results, we found that different predictors are significant for different network concepts. 

For network availability, the results indicate that minimum download goodput (over a user's measurement), number of frequently measured locations, network operator, and device type are significant in determining satisfaction. For network speed, the results indicate that the minimum, median, and most recently measured download goodputs are highly significant in predicting satisfaction. Thus both the peak-end effect and overall integration of download goodput values likely play a part in user evaluation. Furthermore, the predicted probability functions in Figure \ref{predicted_probs_q3} show relatively similar diminishing returns effects on satisfaction for both median and most recent TCP down. 

Additionally, we found very few upload goodput or latency predictors are highly significant for any of the models given the other predictors, thus reinforcing the importance of the download predictors.

We did find several non-network predictors such as the number of frequently measured locations and the temporal burstiness of a user's measurements are significant in several of the models. We postulate that some of these predictors act as proxies for different types of users of Netradar. Interestingly though, we found that our custom measure of device quality was not significant in any of the network related models.

Finally, the overall fits of the result models are only basic thus other factors are also likely at play and further research is required to improve upon the results.

As mentioned, a previous study also found a positive relationship between network availability/speed and user satisfaction \cite{ozer2013}. However, the study was based only on customer surveys and included only download speed in the questionnaire (without upload or latency) and thus lacked the detail of our study. Furthermore, through our model we are able to predict satisfaction in terms of the objectively measured speeds observed by users. Therefore, in terms of novel contributions a large majority of our finding have not been detailed before.

Overall, our results have implications for mobile operators that could utilize such measurement data. For example these operators could target users with predicted low satisfaction (based on their usage of measurement apps) with offers to help prevent churn. Operators could also use the models for service development in terms of predicting satisfaction for speed capped data plans.

In future work, we will expand the usage of the satisfaction questionnaire to include the forthcoming Netradar passive measurement client that tracks network goodput during actual mobile app sessions. Thus we will be able to correlate satisfaction to a more comprehensive view of each user's network performance. We will also look at models to quantify how user network satisfaction could be impacted by, for example, a transition to multi-access mobile devices or the increasing importance of upload in future cloud centric mobile devices.
\section*{Acknowledgment}
This work was supported by the EMERGENT Project (http://emergent.comnet.aalto.fi/).

\bibliography{main}

\begin{thebibliography}{10}
\expandafter\ifx\csname url\endcsname\relax
  \def\url#1{\texttt{#1}}\fi
\expandafter\ifx\csname urlprefix\endcsname\relax\def\urlprefix{URL }\fi
\expandafter\ifx\csname href\endcsname\relax
  \def\href#1#2{#2} \def\path#1{#1}\fi

\bibitem{schatz2014}
R.~Schatz, M.~Fiedler, L.~Skorin-Kapov, Qoe-based network and application
  management, in: S.~M{\"o}ller, A.~Raake (Eds.), Quality of Experience:
  Advanced Concepts, Applications and Methods, T-Labs Series in
  Telecommunication Services, Springer International Publishing, 2014, pp.
  411--426.

\bibitem{mitra2014}
K.~{Mitra}, A.~{Zaslavsky}, C.~{{\AA}hlund}, {QoE Modelling, Measurement and
  Prediction: A Review}, ArXiv e-prints\href {http://arxiv.org/abs/1410.6952}
  {\path{arXiv:1410.6952}}.

\bibitem{alreshoodi2013}
M.~Alreshoodi, J.~Woods, Survey on qoe{\textbackslash}qos correlation models
  for multimedia services, International Journal of Distributed and Parallel
  Systems (IJDPS) 4~(3) (2013) 53--72.

\bibitem{chen2006}
K.-T. Chen, C.-Y. Huang, P.~Huang, C.-L. Lei, Quantifying skype user
  satisfaction, in: Proceedings of the 2006 Conference on Applications,
  Technologies, Architectures, and Protocols for Computer Communications,
  SIGCOMM '06, ACM, 2006, pp. 399--410.

\bibitem{schatz2011}
R.~Schatz, S.~Egger, Vienna surfing: Assessing mobile broadband quality in the
  field, in: Proceedings of the First ACM SIGCOMM Workshop on Measurements Up
  the Stack, W-MUST '11, 2011, pp. 19--24.

\bibitem{casas2015a}
P.~Casas, R.~Schatz, F.~Wamser, M.~Seufert, R.~Irmer, Exploring qoe in cellular
  networks: How much bandwidth do you need for popular smartphone apps?, in:
  Proceedings of the 5th Workshop on All Things Cellular: Operations,
  Applications and Challenges, AllThingsCellular '15, 2015, pp. 13--18.

\bibitem{casas2015b}
P.~Casas, B.~Gardlo, M.~Seufert, F.~Wamser, R.~Schatz, Taming qoe in cellular
  networks: from subjective lab studies to measurements in the field, in:
  Proceedings of the 11th International Conference on Network and Service
  Management (CNSM), 2015, pp. 237--245.

\bibitem{ozer2013}
A.~{\"O}zer, M.~T. Argan, M.~Argan, The effect of mobile service quality
  dimensions on customer satisfaction, Procedia - Social and Behavioral
  Sciences 99 (2013) 428 -- 438, the Proceedings of 9th International Strategic
  Management Conference.

\bibitem{kuo2009}
Y.-F. Kuo, C.-M. Wu, W.-J. Deng, The relationships among service quality,
  perceived value, customer satisfaction, and post-purchase intention in mobile
  value-added services, Computers in Human Behavior 25~(4) (2009) 887 -- 896.

\bibitem{turel2006}
O.~Turel, A.~Serenko, Satisfaction with mobile services in canada: An empirical
  investigation, Telecommunications Policy 30~(5-6) (2006) 314 -- 331.

\bibitem{goel2015}
U.~Goel, M.~Wittie, K.~Claffy, A.~Le, Survey of end-to-end mobile network
  measurement testbeds, tools, and services, Communications Surveys Tutorials,
  IEEE PP~(99) (2015) 1--1.

\bibitem{sonntag2013a}
S.~Sonntag, J.~Manner, L.~Schulte, Netradar - measuring the wireless world, in:
  Modeling Optimization in Mobile, Ad Hoc Wireless Networks (WiOpt), 2013 11th
  International Symposium on, 2013, pp. 29--34.

\bibitem{weiss2014}
B.~Weiss, D.~Guse, S.~M{\"o}ller, A.~Raake, A.~Borowiak, U.~Reiter, Temporal
  development of quality of experience, in: S.~M{\"o}ller, A.~Raake (Eds.),
  Quality of Experience: Advanced Concepts, Applications and Methods, T-Labs
  Series in Telecommunication Services, Springer International Publishing,
  2014, pp. 133--147.

\bibitem{mikkelsen2015}
L.~M. Mikkelsen, N.~B. H{\o}jholt, T.~K. Madsen, Performance evaluation of
  methods for estimating achievable throughput on cellular connections, in:
  S.~Balandin, S.~Andreev, Y.~Koucheryavy (Eds.), Internet of Things, Smart
  Spaces, and Next Generation Networks and Systems, Vol. 9247 of Lecture Notes
  in Computer Science, Springer International Publishing, 2015, pp. 422--435.

\bibitem{callet2013}
S.~M. Patrick Le~Callet, A.~Perkis, Qualinet white paper on definitions of
  quality of experience, Tech. rep., European Network on Quality of Experience
  in Multimedia Systems and Services (COST Action IC 1003), Lausanne,
  Switzerland (03 2013).

\bibitem{kahneman1999}
D.~Kahneman, Objective happiness, in: M.~Kubovy, D.~Kahneman, E.~Diener,
  N.~Schwarz (Eds.), Well-being: The Foundations of Hedonic Psychology, Russel
  Sage, 1999, pp. 3--25.

\bibitem{geng2013}
X.~Geng, Z.~Chen, W.~Lam, Q.~Zheng, Hedonic evaluation over short and long
  retention intervals: The mechanism of the peak-end rule, Journal of
  Behavioral Decision Making 26~(3) (2013) 225--236.

\bibitem{ester1996}
M.~Ester, H.-P. Kriegel, J.~Sander, X.~Xu, A density-based algorithm for
  discovering clusters in large spatial databases with noise., in: Proceedings
  of 2nd International Conference on Knowledge Discovery and Data Mining, 1996,
  pp. 226--231.

\bibitem{goh2008}
K.-I. Goh, A.-L. Barab\'{a}si, Burstiness and memory in complex systems, EPL
  (Europhysics Letters) 81~(4) (2008) 48002.

\bibitem{ender2010}
P.~Ender,
  \href{http://www.ats.ucla.edu/stat/stata/ado/analysis/collin.ado}{Collin
  command - collinearity diagnostics}.
\newline\urlprefix\url{http://www.ats.ucla.edu/stat/stata/ado/analysis/collin.ado}

\bibitem{marquardt1970}
D.~W. Marquardt, Generalized inverses, ridge regression, biased linear
  estimation, and nonlinear estimation, Technometrics 12~(3) (1970) 591--612.

\bibitem{white2011}
I.~R. White, P.~Royston, A.~M. Wood, Multiple imputation using chained
  equations: Issues and guidance for practice, Statistics in Medicine 30~(4)
  (2011) 377--399.

\bibitem{williams2006}
R.~Williams, Generalized ordered logit/partial proportional odds models for
  ordinal dependent variables, Stata Journal 6~(1) (2006) 58--82(25).

\bibitem{sandvine2014}
Sandvine, The global internet phenomena report: 2h 2014, Tech. rep., Sandvine
  (2014).

\bibitem{wac2015}
K.~Wac, G.~Pinar, M.~Gustarini, J.~Marchanoff, Smartphone users mobile networks
  quality provision and volte intend: Six-months field study, in: World of
  Wireless, Mobile and Multimedia Networks (WoWMoM), 2015 IEEE 16th
  International Symposium on a, 2015, pp. 1--9.

\bibitem{mcfadden1975}
T.~Domencich, D.~L. McFadden, Statistical estimation of choice probability
  functions, in: Urban Travel Demand: A Behavioral Analysis, North-Holland
  Publishing Co., 1975, pp. 101--125.

\bibitem{royston2013}
P.~Royston, marginscontplot: Plotting the marginal effects of continuous
  predictors, Stata Journal 13~(3) (2013) 510--527(18).

\bibitem{vesselkov2016}
A.~Vesselkov, H.~H{\"a}mm{\"a}inen,
  \href{http://comnet.aalto.fi/en/midcom-serveattachmentguid-1e5c68b3bf94696c68b11e5acd70192aed101490149/mobile_handset_population_2005-2015.pdf}{Mobile
  handset population in finland 2005-2015}, Tech. rep., Aalto University,
  Department of Communications and Networking (01 2016).
\newline\urlprefix\url{http://comnet.aalto.fi/en/midcom-serveattachmentguid-1e5c68b3bf94696c68b11e5acd70192aed101490149/mobile_handset_population_2005-2015.pdf}

\bibitem{faloutsos2014}
A.~Nikravesh, D.~Choffnes, E.~Katz-Bassett, Z.~Mao, M.~Welsh, Mobile network
  performance from user devices: A longitudinal, multidimensional analysis, in:
  M.~Faloutsos, A.~Kuzmanovic (Eds.), Passive and Active Measurement, Vol. 8362
  of Lecture Notes in Computer Science, Springer International Publishing,
  2014, pp. 12--22.

\bibitem{kobayashi2015}
F.~Kobayashi, G.~Kawaguti, H.~Nojiri, A.~Takahashi, Crowdsourced qoe evaluation
  of mobile device usage with consideration on user dependence, in: Network
  Operations and Management Symposium (APNOMS), 2015 17th Asia-Pacific, 2015,
  pp. 388--391.

\bibitem{hosek2014}
J.~Hosek, P.~Vajsar, L.~Nagy, M.~Ries, O.~Galinina, S.~Andreev, Y.~Koucheryavy,
  Z.~Sulc, P.~Hais, R.~Penizek, Predicting user qoe satisfaction in current
  mobile networks, in: Communications (ICC), 2014 IEEE International Conference
  on, 2014, pp. 1088--1093.

\bibitem{agboma2010}
F.~Agboma, A.~Liotta, Quality of experience management in mobile content
  delivery systems, Telecommunication Systems 49~(1) (2010) 85--98.

\bibitem{menkovski2010}
V.~Menkovski, G.~Exarchakos, A.~Liotta, Machine learning approach for quality
  of experience aware networks, in: 2010 2nd International Conference on
  Intelligent Networking and Collaborative Systems (INCOS), 2010, pp. 461--466.

\bibitem{varela2014}
M.~Varela, L.~Skorin-Kapov, T.~Ebrahimi, Quality of service versus quality of
  experience, in: Quality of Experience, Springer, 2014, pp. 85--96.

\bibitem{reiter2014}
U.~Reiter, K.~Brunnstr{\"o}m, K.~De~Moor, M.-C. Larabi, M.~Pereira,
  A.~Pinheiro, J.~You, A.~Zgank, Factors influencing quality of experience, in:
  Quality of Experience, Springer, 2014, pp. 55--72.

\bibitem{schatz2013}
R.~Schatz, T.~Ho{\ss}feld, L.~Janowski, S.~Egger, From packets to people:
  quality of experience as a new measurement challenge, in: Data traffic
  monitoring and analysis, Springer, 2013, pp. 219--263.

\end{thebibliography}

\newpage
\appendix
\section{Q2 and Q3 Model Coefficients and Cross Correlations}
\label{App:AppendixA}

{
\def\ponepc{$^{\ast\ast\ast}$} \def\onepc{$^{\ast\ast}$} \def\fivepc{$^{\ast}$}
\def\tenpc{$^{\dag}$}
\def\legend{\multicolumn{4}{l}{\footnotesize{Significance levels
:\hspace{1em} $\dag$ : 10\% \hspace{1em}
$\ast$ : 5\% \hspace{1em} $\ast\ast$ : 1\% \normalsize}}}

\begin{table*}[ht!]
\begin{threeparttable}[ht!]\centering
\footnotesize 
\caption{Q2 Partial Proportional Odds Ordinal Logistic Regression Model
\label{Q2_model_coeff}}
\begin{tabular}{l r @{} l c }
\toprule
\multicolumn{1}{c}
{\textbf{Feature}}
 & \multicolumn{2}{c}{\textbf{Coefficient\tnote{a}}}  & \textbf{(Std. Err.)} \\
 \hline
\multicolumn{4}{c}{(1) vs (2,3,4,5)} \\ \hline
Median TCP Down  &  0.012&\fivepc  & (0.006)\\
St. Dev. TCP Down  &  0.015&  & (0.009)\\
Last TCP Down  &  0.004&  & (0.004)\\
Min TCP Down  &  0.088&\onepc  & (0.029)\\
Median TCP Up  &  -0.032&  & (0.026)\\
St. Dev. TCP Up  &  0.058&  & (0.039)\\
Last TCP Up  &  0.006&  & (0.008)\\
Min TCP Up  &  0.001&  & (0.025)\\
Max TCP Up  &  0.001&  & (0.009)\\
Median RTT  &  -0.001&  & (0.000)\\
St. Dev. RTT  &  0.000&  & (0.000)\\
Last RTT  &  0.000&\fivepc  & (0.000)\\
Min RTT  &  0.001&  & (0.001)\\
Max RTT  &  0.000&\fivepc  & (0.000)\\
Number User-Initiated Meas  &  0.000&  & (0.001)\\
Number Meas  &  0.000&  & (0.000)\\
Number Invalid Meas  &  0.000&  & (0.001)\\
Number Invalid User-Initiated Meas  &  -0.006&  & (0.005)\\
Burstiness  &  -0.203&  & (0.225)\\
Mean Device Speed  &  -0.072&\onepc  & (0.026)\\
Number of Frequently Measured Locations &  0.107&\onepc  & (0.040)\\
Operator 1\tnote{b}  & - &  & -\\
Operator 2  &  0.286&\onepc  & (0.088)\\
Operator 3  &  -0.186&  & (0.114)\\
Android\tnote{b}  & - &  & - \\
Windows Phone &  0.245&\fivepc & (0.103)\\
iOS &  -0.109&  & (0.104)\\
High (device quality)\tnote{b} & - & & -\\
Medium (device quality) & 0.247&\tenpc  & (0.143)\\
Low (device quality)  & -0.053&  & (0.096)\\
Smartphone\tnote{b}  & - &  & -\\
Tablet & 0.393&\onepc & (0.125)\\
\hline \multicolumn{4}{c}{(1,2) vs (3,4,5)} \\ \hline
Min TCP Down  &  0.056&\onepc  & (0.018)\\
Median TCP Up  &  0.034&\tenpc  & (0.018)\\
St. Dev. TCP Up  &  -0.008&  & (0.031)\\
Mean Device Speed  &  -0.039&\tenpc  & (0.024)\\
\hline \multicolumn{4}{c}{(1,2,3) vs (4,5)} \\ \hline
Min TCP Down  &  0.041&\ponepc  & (0.012)\\
Median TCP Up  &  0.014&  & (0.013)\\
St. Dev. TCP Up  &  -0.019&  & (0.029)\\
Mean Device Speed  &  0.019&  & (0.021)\\
\hline \multicolumn{4}{c}{(1,2,3,4) vs (5)} \\ \hline
Min TCP Down  &  0.008&  & (0.010)\\
Median TCP Up  &  0.001&  & (0.012)\\
St. Dev. TCP Up  &  0.002&  & (0.029)\\
Mean Device Speed  &  0.022&  & (0.024)\\
\bottomrule
\end{tabular}
\begin{tablenotes}
      \small
      \item[a] $\dag$ : 10\%, $\ast$ : 5\%, $\ast\ast$ : 1\%, $\ast{\ast}\ast$ : 0.1\%
      \item[b] The base category of a categorical variable.
\end{tablenotes}
\end{threeparttable}
\end{table*}
}

{
\def\ponepc{$^{\ast\ast\ast}$} \def\onepc{$^{\ast\ast}$} \def\fivepc{$^{\ast}$}
\def\tenpc{$^{\dag}$}
\def\legend{\multicolumn{4}{l}{\footnotesize{Significance levels
:\hspace{1em} $\dag$ : 10\% \hspace{1em}
$\ast$ : 5\% \hspace{1em} $\ast\ast$ : 1\% \normalsize}}}
\begin{table*}[ht!]
\begin{threeparttable}[ht!]\centering
\footnotesize
 \caption{Q3 Partial Proportional Odds Ordinal Logistic Regression Model
\label{Q3_model_coeff}}
\begin{tabular}{l r @{} l c }
\toprule
\multicolumn{1}{c}
{\textbf{Feature}}
 & \multicolumn{2}{c}{\textbf{Coefficient\tnote{a}}}  & \textbf{(Std. Err.)} \\
\hline \multicolumn{4}{c}{(1) vs (2,3,4,5)} \\ \hline
Median TCP Down & 0.019&\onepc  & (0.006)\\
St. Dev. TCP Down & 0.001&  & (0.009)\\
Last TCP Down   &  0.017&\ponepc  & (0.004)\\
Min TCP Down   &  0.075&\onepc  & (0.025)\\
Median TCP Up  &  0.017&  & (0.011)\\
St. Dev. TCP Up  &  0.021&  & (0.027)\\
Last TCP Up  &  0.001&  & (0.007)\\
Min TCP Up  &  -0.008&  & (0.025)\\
Max TCP Up  &  -0.003&  & (0.008)\\
Median RTT  &  -0.001&  & (0.000)\\
St. Dev. RTT  &  0.000&  & (0.000)\\
Last RTT  &  0.000&  & (0.000)\\
Min RTT  &  0.000&  & (0.001)\\
Max RTT  &  0.000&  & (0.000)\\
Number User-Init Meas &  -0.002&  & (0.001)\\
Number Meas  &  0.000&  & (0.000)\\
Number Invalid Meas  &  0.000&  & (0.001)\\
Number Invalid User-Init Meas & -0.013&\fivepc  & (0.006)\\
Burstiness  &  -0.384&\tenpc  & (0.220)\\
Mean Device Speed  &  0.025&  & (0.018)\\
Number of Frequently Measured Locations &  0.089&\fivepc  & (0.036)\\
Operator 1\tnote{b} & - &  & -\\
Operator 2  &  0.118 &  & (0.087)\\
Operator 3  & -0.033 &  & (0.113)\\
Android\tnote{b} & - &  & - \\
Windows Phone  &  0.225&\fivepc & (0.102)\\
iOS & -0.262&\tenpc  & (0.156)\\
High (device quality)\tnote{b} & - & & -\\
Medium (device quality) & 0.163&\tenpc  & (0.140)\\
Low (device quality)  & -0.151&  & (0.095)\\
Smartphone\tnote{b}  & - &  & -\\
Tablet & 0.091& & (0.121)\\
\hline \multicolumn{4}{c}{(1,2) vs (3,4,5)} \\ \hline
Min TCP Down  &  0.050&\ponepc  & (0.014)\\
iOS  &  -0.148&  & (0.123)\\
\hline \multicolumn{4}{c}{(1,2,3) vs (4,5)} \\ \hline
Min TCP Down  &  0.019&\tenpc  & (0.011)\\
iOS   &  -0.122&  & (0.116)\\
\hline \multicolumn{4}{c}{(1,2,3,4) vs (5)} \\ \hline
Min TCP Down  &  -0.006&  & (0.010)\\
iOS  &  0.173&  & (0.140)\\
\bottomrule
\end{tabular}
\begin{tablenotes}
      \small
      \item[a] $\dag$ : 10\%, $\ast$ : 5\%, $\ast\ast$ : 1\%, $\ast{\ast}\ast$ : 0.1\%
      \item[b] The base category of a categorical variable.
\end{tablenotes}
\end{threeparttable}
\end{table*}
}

\begin{sidewaystable*}
\begin{threeparttable}
\small
\caption{\label{numeric_feature_correlations} Measurement Features Cross-Correlation Coefficients\tnote{a}}\centering\medskip
\tabcolsep=0.11cm
\begin{tabular}{|l|l|l|l|l|l|l|l|l|l|l|l|l|l|l|l|l|l|l|l|l|l|l|l|l|l|}\hline  
 & a & b  & c  & d  & e  & f  & g  & h  & i  & j  & k  & l  & m  & n  & o  & p  & q  & r  & s  & t  & u  & v  & w  & x  & y  \\ \hline  
a & 1.00 \\ \hline 
b & 0.83 & 1.00 \\ \hline 
c & 0.64 & 0.50 & 1.00 \\ \hline 
d & 0.54 & 0.51 & 0.41 & 1.00 \\ \hline 
e & 0.34 & 0.37 & 0.09 & 0.31 & 1.00 \\ \hline 
f & 0.71 & 0.58 & 0.78 & 0.48 & 0.17 & 1.00 \\ \hline 
g & 0.67 & 0.61 & 0.54 & 0.43 & 0.28 & 0.59 & 1.00 \\ \hline 
h & 0.56 & 0.59 & 0.38 & 0.39 & 0.31 & 0.44 & 0.73 & 1.00 \\ \hline 
i & 0.57 & 0.48 & 0.64 & 0.38 & 0.12 & 0.62 & 0.69 & 0.45 & 1.00 \\ \hline 
j & 0.38 & 0.37 & 0.29 & 0.55 & 0.24 & 0.34 & 0.46 & 0.46 & 0.35 & 1.00 \\ \hline 
k & 0.18 & 0.22 & -0.02 & 0.16 & 0.51 & 0.03 & 0.21 & 0.30 & 0.00 & 0.24 & 1.00 \\ \hline 
l & 0.58 & 0.49 & 0.61 & 0.40 & 0.13 & 0.67 & 0.71 & 0.49 & 0.81 & 0.39 & 0.04 & 1.00 \\ \hline 
m & -0.38 & -0.39 & -0.21 & -0.29 & -0.42 & -0.27 & -0.36 & -0.39 & -0.23 & -0.26 & -0.33 & -0.26 & 1.00 \\ \hline 
n & -0.43 & -0.43 & -0.31 & -0.30 & -0.26 & -0.34 & -0.40 & -0.38 & -0.32 & -0.24 & -0.15 & -0.33 & 0.64 & 1.00 \\ \hline 
o & -0.26 & -0.28 & -0.06 & -0.21 & -0.48 & -0.12 & -0.25 & -0.31 & -0.09 & -0.21 & -0.45 & -0.12 & 0.65 & 0.31 & 1.00 \\ \hline 
p & -0.32 & -0.31 & -0.24 & -0.42 & -0.22 & -0.28 & -0.30 & -0.27 & -0.25 & -0.38 & -0.13 & -0.27 & 0.46 & 0.56 & 0.26 & 1.00 \\ \hline 
q & -0.33 & -0.29 & -0.32 & -0.23 & -0.11 & -0.35 & -0.30 & -0.23 & -0.32 & -0.16 & 0.00 & -0.33 & 0.42 & 0.62 & 0.13 & 0.53 & 1.00 \\ \hline 
r & -0.24 & -0.27 & -0.06 & -0.21 & -0.52 & -0.10 & -0.23 & -0.28 & -0.09 & -0.20 & -0.48 & -0.09 & 0.65 & 0.36 & 0.79 & 0.32 & 0.18 & 1.00 \\ \hline 
s & -0.04 & -0.05 & -0.01 & -0.01 & -0.11 & 0.02 & -0.02 & -0.04 & -0.01 & -0.01 & -0.12 & 0.03 & 0.07 & 0.04 & 0.09 & 0.03 & -0.02 & 0.12 & 1.00 \\ \hline 
t & 0.06 & 0.04 & 0.12 & 0.04 & -0.25 & 0.17 & 0.08 & 0.05 & 0.11 & 0.04 & -0.25 & 0.18 & 0.06 & -0.03 & 0.16 & -0.04 & -0.13 & 0.24 & 0.25 & 1.00 \\ \hline 
u & 0.12 & 0.08 & 0.20 & 0.07 & -0.12 & 0.20 & 0.13 & 0.09 & 0.19 & 0.06 & -0.15 & 0.21 & -0.01 & -0.06 & 0.08 & -0.04 & -0.13 & 0.11 & 0.10 & 0.23 & 1.00 \\ \hline 
v & 0.06 & 0.03 & 0.15 & 0.04 & -0.30 & 0.21 & 0.08 & 0.03 & 0.14 & 0.04 & -0.33 & 0.22 & 0.08 & -0.03 & 0.22 & -0.02 & -0.15 & 0.31 & 0.27 & 0.59 & 0.27 & 1.00 \\ \hline 
w & 0.04 & 0.01 & 0.13 & 0.02 & -0.29 & 0.17 & 0.05 & 0.01 & 0.11 & 0.02 & -0.31 & 0.17 & 0.11 & -0.01 & 0.26 & -0.01 & -0.11 & 0.30 & 0.24 & 0.50 & 0.24 & 0.82 & 1.00 \\ \hline 
x & -0.13 & -0.15 & -0.01 & -0.11 & -0.33 & -0.03 & -0.13 & -0.16 & -0.04 & -0.10 & -0.31 & -0.02 & 0.23 & 0.12 & 0.31 & 0.08 & 0.00 & 0.31 & 0.18 & 0.02 & 0.15 & 0.36 & 0.43 & 1.00 \\ \hline 
y & -0.14 & -0.15 & -0.02 & -0.12 & -0.35 & -0.02 & -0.13 & -0.17 & -0.04 & -0.11 & -0.33 & -0.02 & 0.23 & 0.13 & 0.29 & 0.10 & 0.00 & 0.32 & 0.18 & 0.00 & 0.15 & 0.39 & 0.35 & 0.90 & 1.00 \\ \hline  
\end{tabular}
  \begin{tablenotes}
      \small
      \item[a] a - mean TCP down, b - median TCP down, c - std TCP down, d - last TCP down, e - min TCP down, f - max TCP down, g - mean TCP up, h - median TCP up, i - std TCP up, j - last TCP up, k - min TCP up, l - max TCP up, m - mean RTT, n - median RTT, o - std RTT, p - last RTT, q - min RTT, r - max RTT, s - burstiness, t - number of frequently measured locations, u - mean device speed, v - number user-initiated meas, w - number meas, x - number invalid meas, y - number invalid user-initiated meas
  \end{tablenotes}
\end{threeparttable}
\end{sidewaystable*}

\begin{sidewaystable*}
\begin{threeparttable}
\caption{\label{numeric_feature_significances} Measurement Features Cross-Correlation Significance Levels\tnote{a}\tnote{b}}\centering\medskip
\tabcolsep=0.11cm
\begin{tabular}{|l|l|l|l|l|l|l|l|l|l|l|l|l|l|l|l|l|l|l|l|l|l|l|l|l|l|}\hline  
 & a & b  & c  & d  & e  & f  & g  & h  & i  & j  & k  & l  & m  & n  & o  & p  & q  & r  & s  & t  & u  & v  & w  & x  & y  \\ \hline
a &   - \\ \hline 
b &   *** &   - \\ \hline 
c &   *** &   *** &   - \\ \hline 
d &   *** &   *** &   *** &   - \\ \hline 
e &   *** &   *** &   *** &   *** &   - \\ \hline 
f &   *** &   *** &   *** &   *** &   *** &   - \\ \hline 
g &   *** &   *** &   *** &   *** &   *** &   *** &   - \\ \hline 
h &   *** &   *** &   *** &   *** &   *** &   *** &   *** &   - \\ \hline 
i &   *** &   *** &   *** &   *** &   *** &   *** &   *** &   *** &   - \\ \hline 
j &   *** &   *** &   *** &   *** &   *** &   *** &   *** &   *** &   *** &   - \\ \hline 
k &   *** &   *** &       &   *** &   *** &       &   *** &   *** &       &   *** &   - \\ \hline 
l &   *** &   *** &   *** &   *** &   *** &   *** &   *** &   *** &   *** &   *** &       &   - \\ \hline 
m &   *** &   *** &   *** &   *** &   *** &   *** &   *** &   *** &   *** &   *** &   *** &   *** &   - \\ \hline 
n &   *** &   *** &   *** &   *** &   *** &   *** &   *** &   *** &   *** &   *** &   *** &   *** &   *** &   - \\ \hline 
o &   *** &   *** &   **  &   *** &   *** &   *** &   *** &   *** &   *** &   *** &   *** &   *** &   *** &   *** &   - \\ \hline 
p &   *** &   *** &   *** &   *** &   *** &   *** &   *** &   *** &   *** &   *** &   *** &   *** &   *** &   *** &   *** &   - \\ \hline 
q &   *** &   *** &   *** &   *** &   *** &   *** &   *** &   *** &   *** &   *** &       &   *** &   *** &   *** &   *** &   *** &   - \\ \hline 
r &   *** &   *** &   **  &   *** &   *** &   *** &   *** &   *** &   *** &   *** &   *** &   *** &   *** &   *** &   *** &   *** &   *** &   - \\ \hline 
s &       &       &       &       &   *** &       &       &       &       &       &   *** &       &   *** &       &   *** &       &       &   *** &   - \\ \hline 
t &       &       &   *** &       &   *** &   *** &   *** &       &   *** &       &   *** &   *** &       &       &   *** &       &   *** &   *** &   *** &   - \\ \hline 
u &   *** &   *** &   *** &   **  &   *** &   *** &   *** &   *** &   *** &   *   &   *** &   *** &       &   *   &   *** &       &   *** &   *** &   *** &   *** &      - \\ \hline 
v &   *   &       &   *** &       &   *** &   *** &   *** &       &   *** &       &   *** &   *** &   *** &       &   *** &       &   *** &   *** &   *** &   *** &  *** &   - \\ \hline 
w &       &       &   *** &       &   *** &   *** &       &       &   *** &       &   *** &   *** &   *** &       &   *** &       &   *** &   *** &   *** &   *** &  *** &   *** &   - \\ \hline 
x &   *** &   *** &       &   *** &   *** &       &   *** &   *** &       &   *** &   *** &       &   *** &   *** &   *** &   *** &       &   *** &   *** &   *** &  *** &   *** &   *** &   - \\ \hline 
y &   *** &   *** &       &   *** &   *** &       &   *** &   *** &       &   *** &   *** &       &   *** &   *** &   *** &   *** &       &   *** &   *** &   *** &    *** &   *** &   *** &   *** &   - \\ \hline 
\end{tabular}
  \begin{tablenotes}
      \small
      \item[a] a - mean TCP down, b - median TCP down, c - std TCP down, d - last TCP down, e - min TCP down, f - max TCP down, g - mean TCP up, h - median TCP up, i - std TCP up, j - last TCP up, k - min TCP up, l - max TCP up, m - mean RTT, n - median RTT, o - std RTT, p - last RTT, q - min RTT, r - max RTT, s - burstiness, t - number frequently measured locations, u - mean device speed, v - number user-initiated meas, w - number meas, x - number invalid meas, y - number invalid user-initiated meas
      \item[b] $\dag$ : 10\%, $\ast$ : 5\%, $\ast\ast$ : 1\%, $\ast{\ast}\ast$ : 0.1\%
  \end{tablenotes}
\end{threeparttable}
\end{sidewaystable*}
\end{document}